\documentclass[]{article}
\usepackage{amsmath}
\usepackage{amsfonts}
\usepackage{amssymb}
\usepackage{todonotes}
\usepackage[skip=2pt]{caption}
\usepackage{authblk}
\usepackage{float}

\newcommand\blfootnote[1]{%
  \begingroup
  \renewcommand\thefootnote{}\footnote{#1}%
  \addtocounter{footnote}{-1}%
  \endgroup
}

\title{An Enhanced Initial Margin Methodology to Manage Warehoused Credit Risk}

\author[1]{Lucia Cipolina-Kun}
\author[2]{Ignacio Ruiz}
\author[3]{Mariano Zeron-Medina Laris}

\affil[1]{Morgan Stanley}
\affil[2]{MoCaX Intelligence by iRuiz Technologies.  i.ruiz@iruiztechnologies.com}
\affil[3]{MoCaX Intelligence by iRuiz Technologies. m.zeron@iruiztechnologies.com}

\begin{document}
\maketitle
\begin{abstract}

\blfootnote{The authors would like to thank Andrew S.Dickinson, Martin Dahlgren and Wujiang Lou}

\blfootnote{The opinions expressed here are those of the authors, and do not necessarily represent the view of their employers}

\noindent The use of CVA to cover credit risk is widely spread, but has its limitations. Namely, dealers face the problem of the illiquidity of instruments used for hedging it, hence forced to warehouse credit risk. As a result, dealers tend to offer a limited OTC derivatives market to highly risky counterparties. Consequently, those highly risky entities rarely have access to hedging services precisely when they need them most.

\medskip
\noindent In this paper we propose a method to overcome this limitation. We propose to extend the CVA risk-neutral framework to compute an initial margin (IM) specific to each counterparty, which depends on the credit quality of the entity at stake, transforming the effective credit rating of a given netting set to AAA, regardless of the credit rating of the counterparty.

\medskip
\noindent The proposed methodology is fully compliant with the risk-neutral pricing framework, enables improved risk management for dealers and, subsequently, a route for troubled institutions to access the derivative markets.

\medskip
\noindent By transforming CVA requirement into IM ones, as proposed in this paper, an institution could rely on the existing mechanisms for posting and calling of IM, hence ensuring the operational viability of this new form of managing warehoused risk. The main difference with the currently standard framework is the creation of a Specific Initial Margin, that depends in the credit rating of the counterparty and the characteristics of the netting set in question. In this paper we propose a methodology for such transformation in a sound manner, and hence this method overcomes some of the limitations of the CVA framework.

\medskip
\noindent Using a range of Swaps and Swaptions, we present realistic numerical examples of CVA values under margining, IM values as computed nowadays, as well as the new amounts of Specific Initial Margin computed under the framework introduced in this paper. In our opinion, transition to this enhance frameworks could be smooth, and in the interest of all parties, including broker-dealers and derivative users.
\medskip

\end{abstract}

\section{Introduction}
\label {sec:intro}

\noindent 
In response to the $2007-08$ crisis, Regulators proposed a series of measures with the aim of decreasing the interbank counterparty credit risk that, in a domino-like effect, brought the interbank market to a halt during the credit crunch. The new rules are motivated by the clearing and margining mechanisms long adopted by Central Clearing houses which have been perceived to be effective in curtailing the contagion effects of defaulting counterparties.

\medskip
\noindent 
Two key regulations have been passed. First, regulators want derivatives dealers to clear as many trades as possible through clearing houses. These houses require clearing members to post a set of collateral amounts composed of at least variation margin (VM), initial margin (IM) and a default fund contribution. Second, as proposed by the Basel Committee of Banking Supervision in \cite{BCBS}, uncleared trades between the vast majority of financial institutions and large corporates must be subject to variation and initial margining, both posted daily.

\medskip
\noindent 
Financial institutions have been posting variation margin for bilateral trades in the past, however posting bilateral IM is new to the industry. The rationale under the new bilateral IM requirement is to use IM as a buffer against the gap risk of netting sets previously covered by VM only\footnote{By gap risk we mean the adverse change in the netting set value after the counterparty’s default. It is also called “close-out risk”.}. If a netting set is only collateralized with VM, in a default event, the surviving party still faces the risk arising from the market movements until the defaulted trades are either wound down or re-hedged. Under the new IM requirement for bilateral trades, the IM amount is posted into a segregated account, therefore, the surviving entity can take the IM posted by the defaulted counterparty to compensate for the losses it may incur during the close-out period.

\medskip
\noindent 
The basis of the modelling of bilateral IM and its settling rules were jointly proposed in $2015$ by the Basel Committee of Banking Supervision (BCBS) and the International Organisation of Securities Commissions (IOSCO) \cite{BCBS}. There are two regulatory schemes used to compute bilateral IM of a netting set. The first is a simple “Schedule-based” approach that calculates a trade-level IM amount based on a percentage of its notional basis. Under this scheme, no IM netting is allowed, therefore this method is generally avoided by financial institutions as it tends to be very costly\footnote{Exceptions may include institutions with highly directional netting sets such as pension funds.}. The second is an advanced “Model-based” approach. Under this scheme, IM is determined as the $99$-percentile loss of the netting set over the margin period of risk (MPoR), computed under stressed market conditions.

\medskip
\noindent 
Given that IM under the Model-based approach is substantially lower than its Schedule-based equivalent as soon as netting effects are moderate, most institutions are implementing Model-based margining for their books of uncleared trades.

\medskip
\noindent 
The interbank association ISDA has put forward the risk-based Standard Initial Margin Model (SIMM™) \cite{ISDA-SIMM} with the aim of standardising the model-based IM that each institution calculates and hence minimise the chances of IM disputes. ISDA’s SIMM model has become very relevant as it has been adopted widely across the industry and approved by regulators across many jurisdictions\footnote{The mandatory implementation of SIMM takes place between 2016 and 2020, and aims to incorporate all relevant financial institutions and corporates. The largest financial institutions began exchanging bilateral initial and variation margin from September 1 (2016), under rules that took effect in the US, Japan and Canada.}.

\medskip
\noindent 
This is the form of IM we assume for the rest of the paper. It must be noted, however, that the Specific Initial Margin we propose can be defined for IM computed under any scheme.

\subsection{The Framework}

\medskip
\noindent 
Margining as described above significantly reduces counterparty credit risk but does not cover it in its entirety. As mentioned before, bilateral IM models should cover $99\%$ of netting set losses during the MPoR. The fraction of the risk left uncovered can be measured via the Credit Value Adjustment (CVA). CVA is the risk-neutral price of the counterparty credit risk embedded in a netting set. In theory, this price equals the cost of hedging the default risk of the netting set in question. However, after the introduction of mandatory VM and IM, the high amount of collateralization results in CVA amounts that are virtually negligible compared to the monetary size of the transaction. Works by Andersen et al. \cite{Andersen-Exposure} and Gregory \cite{Gregory} estimate that the introduction of IM will reduce the expected exposure by approximately two orders of magnitude. Our numerical results support this estimate as can be seen in table \ref{tab:table1} below.

\medskip
\noindent 
As a result of this reduction in the expected exposure, the CVA in a collateralized transaction is often deemed too small to be exchanged. Even if it is exchanged, this risk cannot always be hedged in practice. For example, the Credit Default Swap market used for hedging often does not have the needed depth. This restricts the trading options of a dealer as it is not compensated for the small but still present counterparty credit risk, according to the credit rating of the counterparty.


\medskip
\noindent To exemplify the point, say we are a derivatives dealer with two clients, both subject to VM and IM under the CSA terms. Let us say that, today, there is only one significant difference between the clients: Counterparty $1$ is an entity with a strong credit standing (say AAA), while Counterparty $2$ is poorly rated (say CCC). If the dealer perceives the credit risk coming from each counterparty to be roughly equal because CVA is so small that it is not charged, she will be virtually equally inclined to trade with either counterparty. This is financially unsound as Counterparty $1$ is a better entity than  Counterparty $2$. It is also suboptimal from the risk management standpoint because the added riskiness of Counterparty $2$ is not taken into account, leading to risk-skewed balance sheets. If, on the contrary, CVA is charged, however large or small it may be, the market will typically not offer Credit Default Swaps for it, hence its credit risk cannot be hedged, and the derivative dealer has no other option than warehousing the risk.

\medskip
\noindent Warehousing credit risk from poorly rated counterparties is not something desired by dealers for obvious reasons. Hence, dealers do not tend to offer them derivative transactions. Often this happens when these hedging services are most needed.

\medskip
\noindent We put forward the following question: is there a way for a derivatives dealer to transform a CCC risk into a AAA risk, despite the credit hedging limitations of the market?

\medskip
\noindent The answer to the question above is yes. In this paper we propose a methodology to achieve this by defining what we call \textit{Specific Initial Margin}, obtained by requesting an additional IM amount on top of the current IM as given by the SIMM model, that depends on the credit quality of the counterparty. This means that each counterparty will post an IM amount according to its credit rating\footnote{Note that under the current ISDA rules, an entity is allowed to request an IM higher than SIMM to a counterparty. Therefore, there are no legal constraints that apply to our method.}.

\medskip
\noindent The proposed calculation process involves reducing the CVA of any given counterparty down to that of the strongest counterparty (e.g. AAA) by increasing the IM demanded to the counterparty in question. By doing this, the Counterparty Credit Risk is reduced to that of a strong entity (e.g. AAA) even when there is no market to hedge it out, because the dealer has an extra buffer protection in the form of IM posted in a segregated account. In this way, we are transforming the credit risk of a poorly rated counterparty into funding cost for the counterparty. Of course, if the counterparty is poorly rated, its funding cost will be high, but how to deal with that problem is beyond the scope of this paper. In this paper we propose a practical way for a dealer to offer derivative products to a counterparty, that has no credit-hedging market, without having to warehouse the risk.

\medskip
\noindent The paper is organised as follows. Section \ref{sec:IM in the Context of Risk Neutral Pricing} describes IM within the context of the risk neutral pricing framework and how our proposal fits within this context. Section \ref{sec:Specific Initial Margin} presents the details of how we compute the Specific Initial Margin of each counterparty. Section \ref{sec:Numerical Tests} presents computed values of Specific Initial Margin for a range of counterparties and a range of Swaps and Swaptions (of varying maturities). In Section \ref{sec:Conclusion} we end with a brief conclusion.

\section{IM in the Context of Risk Neutral Pricing}
\label {sec:IM in the Context of Risk Neutral Pricing}

\medskip
\noindent CVA is the risk-neutral price of the counterparty credit risk embedded in a netting set. Following the replication approach \cite{Burgard}, this price must be equal to the cost of hedging the default risk of the netting set at stake. Risk-neutral pricing theory assumes that all risks can be hedged. When done properly we end up with a risk-free netting set. In this way, a broker dealer selling a derivative should make the risk-free rate of return, plus the margin it charges its clients for the services provided \footnote{By services it is meant the provision of OTC derivatives for the needed hedging purposes of the client.}.

\medskip
\noindent Given the high level of margin required (i.e. at a $99\%$ level), collateralizing the exposure with VM and IM yields a quasi-default-risk-free netting set. As a result, CVA is very small compared to the monetary size of the transaction. The chart below shows the disparity between CVA values and the Notionals of the corresponding trades, before and after collateralization. Given the small values of Collateralised CVA, it is often not exchanged.

\medskip
\begin {table}[!htb]
\caption {CVA for different collateralization schemes for an at-the-money swap.} \label{tab:table1} 
\begin{center}
\begin{tabular} {p{1cm} p{2.1cm}p{1.5cm}p{2.5cm}p{2.5cm}p{2.5cm}}
\hline 
Rating & Prob default & Notional & Uncollateralised & Collateralised & Collateralised  \\ 
       &  (basis points)  & &  CVA & CVA (VM)  & CVA (VM-IM) \\ 
\hline 
AAA & 1 & $\$$1,000,000  & $\$$8,238 & $\$$98 & $\$$0.02 \\ 
\hline 
CCC & 2682  & $\$$1,000,000 & $\$$570,198 & $\$$9,152 & $\$$0.25 \\ 
\hline 
\end{tabular} 
   \caption*{Figures in US dollars}
\end{center}
\end {table}

\newpage
\bigskip
\noindent As CVA with VM and IM becomes very small for all counterparties, there is no practical way for a dealer to measure the difference in counterparty credit risk \footnote{Counterparty credit risk that is small but still present.} between counterparties with different ratings, with the subsequent discussed limitations this leads to. To solve this problem, we propose adjusting the IM amount requested to each counterparty. hola

\medskip
\noindent As a motivating example, say we have the same situation as the one presented in Section \ref{sec:intro}: a dealer facing two counterparties; Counterparty $1$ is AAA rated with a CVA for a given netting set of, say, $\$1$, while Counterparty $2$ is CCC rated with a CVA for the same netting set of, say, $\$10$. The difference in premiums reflects the difference in counterparty credit risk between the two of them. Assume the CVA is too small to be exchanged and/or that the dealer cannot hedge the default risk with Counterparty $2$. What we propose is for the dealer to ask the Counterparty $2$ (the CCC one) for extra IM so that its counterparty risk is decreased to the level of Counterparty $1$ (the AAA one), making them equivalent from a credit-risk worthiness. The natural question is: how much IM should the dealer demand from Counterparty $2$?

\medskip
\noindent We solve this problem by increasing the IM requirements of Counterparty 2 so that its CVA is equal to the CVA of Counterparty $1$ (remember we assume identical netting sets in this illustrative example). By doing so, the respective netting sets of both counterparties end up being risk equivalent and of AAA quality. Under the newly proposed scenario, the dealer is free to charge both Counterparty $1$ and Counterparty $2$ a CVA of $1$. By posting an extra IM amount, we have effectively converted the netting set of Counterparty $2$ into a AAA-equivalent one. Counterparty $2$ will have to pay an extra funding cost and face liquidity risk on the extra IM, but that issue is beyond the scope of this piece of work. Details of how to compute the extra IM needed for a given counterparty and trade in question are presented in Section \ref{sec:Specific Initial Margin}.
\medskip

\noindent The rationale of the method is to reduce the problematic CVA as much as possible: down to the CVA of the highest rated counterparty, by adjusting the IM according to the credit rating of the counterparty, thus compensating the dealer for the credit risk inherent in the counterparty.
\medskip

\noindent It must be noted that we are not challenging the risk-neutral pricing theory but complementing it; compensating for the assumption that all risks can be hedged, when they sometimes cannot. Within this framework, we implicitly acknowledge that default risk cannot be hedged and devise an alternative strategy, via margining, to make the netting set quasi-default-risk-free.

\section{Specific Initial Margin}
\label {sec:Specific Initial Margin}

\medskip
\noindent In this section we define \textit{Specific Initial Margin} for any given counterparty. As it has already been mentioned, we do this by reducing the CVA of the counterparty in question to that of a AAA-quality one by increasing the IM demanded of it. Assuming the lowest CVA corresponds to the counterparty with the highest credit rating, the minimum IM any counterparty is asked for is the amount of IM that is currently computed in the industry (e.g. SIMM™). We decompose $IM^j_{total}$, the total initial margin demanded of counterparty $j$, into two components

\begin{equation}
\label{eq:1}
IM^j_{\text{specific}} = IM_{\text{general}} + IM^j_{\text{add-on}} 
\end{equation}

\medskip
\noindent Where $IM_{\text{general}}$ accounts for the Gap risk without any consideration of the credit quality of the counterparty (e.g. SIMM™), and $IM^j_{\text{add-on}} $ is the add-on that accounts for the portion of $IM^j_{\text{specific}}$ that is specific to the counterparty. For reasons that will become clear, we express $IM^j_{\text{add-on}}$ as a proportion of $IM_{\text{general}}$ giving us

\begin{equation}
\label{eq:IMtotal}
IM^j_{\text{specific}} = IM_{\text{general}} + \alpha^j IM_{\text{general}} 
\end{equation}

\noindent where $\alpha^j$ is a real value that depends on the default probability of counterparty $j$.

\medskip
\noindent Take the definition of CVA presented in \cite{Brigo}. For counterparty $i$ this is

\begin{equation}
\label{eq:CVA1}
CVA_0^i = \mathbb{E}\left[ \int_{0}^{T} \left( V_{t}-VM_{t}-IM_{t}  \right)^+ LGD^i_t PD^i_t DF_t dt     \right]  
\end{equation}

\noindent where $V_{t}$, $VM_{t}$ and $IM_{t}$ are, respectively, the value of the netting set, the VM posted/received, and the IM available to the Bank at time $t$. The term $()^+$ symbolizes zero-flooring, $ LGD^i_t$ the Loss Given Default of the counterparty in question at time t; $PD^i_t$ the counterparty marginal default probability at time t; $DF_t$  the risky discount factor at time $t$, and $T$ the time to maturity of the netting set \footnote{By marginal default probability at time $t$ we mean the default probability between $t$ and $t+dt$, given that the counterparty has survived up to time $t$.}.   
\medskip

\noindent Assume Counterparty $i$ to be the counterparty with the best credit rating available in the market and Counterparty $j$ any other counterparty. Note that in Equation \ref{eq:CVA1}, the exposure part of the equation is the same for both counterparties while the term that distinguishes counterparties is the marginal default probability. The LGD can be assumed to be the same for both counterparties given the context of this calculation\footnote{As the use of this framework is to compute the extra IM needed to reduce the counterparty credit risk of a given entity to that of a AAA, we can assume, without loss of generalization, the LGD to be constant and the same for different counterparties.}. In Equation \ref{eq:CVA1}, until now, $ IM_t$  corresponds to the value of $IM_{\text{general}}$ from Equation \ref{eq:IMtotal} at each time point $t$ in the future, which yields, different CVA values for different counterparties, everything else being the same. To make them equal, we replace $IM_t$ by  $ IM_t  + \alpha^j IM_{t}$ from Equation \ref{eq:IMtotal}, where $IM_t$ will typically be given by the SIMM. 

\begin{equation}
\label{eq:CVA2}
CVA_0^j = \mathbb{E}\left[ \int_{0}^{T} \left( V_{t}-VM_{t}-IM_{t}-\alpha^j IM_{t}  \right)^+ LGD^j_t PD^j_t DF_t dt     \right]  
\end{equation}

\medskip
\noindent The objective is to compute the value $ \alpha^j$ in Equation \ref{eq:CVA2} corresponding to Counterparty $j$  that makes the CVA of Counterparty $j$ equal to the CVA of Counterparty $i$.

\medskip
\noindent We compute $\alpha^j$ as follows. First, the CVA for Counterparty $i$ is computed using equation \ref{eq:CVA1} by running a Monte Carlo simulation. Note that this involves simulating the risk factors that drive the value of the netting set and computing the exposure (i.e. $V_{t}-VM_{t}-IM_{t}$), at each time step $t$ and Monte Carlo path. Second, within the same Monte Carlo simulation, we numerically solve for $\alpha^j$ in the following equation

\begin{multline} \label{eq:CVA3}
\mathbb{E}\left[ \int_{0}^{T} \left( V_{t}-VM_{t}-IM_{t}  \right)^+ LGD_t PD^i_t DF_t dt  \right]\\
 - \mathbb{E}\left[ \int_{0}^{T} \left( V_{t}-VM_{t}-IM_{t}-\alpha^j IM_{t}  \right)^+ LGD_t PD^j_t DF_t dt \right] =0 
\end{multline}
\medskip

\medskip
\noindent Note that we assume the value  $\alpha^j$ used in the Monte Carlo simulation is constant. That is, it is independent of the node defined by each path and time point within the Monte Carlo simulation. Therefore, the add-on Initial Margin ($\alpha^jIM_{t}$) at each node of the simulation is a constant proportion of the initial margin $IM_{t}$\footnote{Strictly speaking, in a full-world simulation, the parameter $\alpha^j$ should depend on the simulated credit worthiness of the counterparty at each Monte Carlo node. However, this refinement is expected to lead only to second-order adjustments in today’s value of $\alpha^j$, so we have decided to leave it constant throughout the Monte Carlo simulation. Further research could shed light onto the validity of this hypothesis.}.

\medskip
\noindent Lastly, the value $\alpha^j$ that solves Equation \ref{eq:CVA3} is used in Equation \ref{eq:1} to determine $IM^j_{\text{add-on}}$ for Counterparty $2$ and compute today’s  $IM^j_{\text{specific}}$ .

\medskip
\noindent As stated in Section \ref{sec:intro}, under this new framework, the specific initial margin requested (i.e. $I{M^j}_{\text{specific}}$), depends on the difference in credit ratings between counterparties $i$ and $j$. Specifically, from Equation \ref{eq:CVA3}, we can see that the lower the rating of a counterparty, the higher  $I{M^j}_{\text{specific}}$ will tend to be. By reducing the CVA to virtually risk-free levels (i.e. those of a AAA-rated counterparty), the IM now not just covers $99\%$ of possible netting set losses during the MPoR, but also compensates for the counterparty credit risk that CVA is meant to cover.

\medskip
\noindent Equivalently, one can think of $IM^j_{\text{specific}}$ as an IM value that has been calibrated at a higher percentile, where this percentile depends on the counterparty’s credit ratings.

\section{Formulaic Approach}
\label {sec:Formulas}

\medskip
\noindent In the following formulas we take the difference between Equation \ref{eq:CVA2} (CVA computed with $IM_{\text{specific}}$) and Equation \ref{eq:CVA1} (CVA computed with $IM_{\text{general}}$), to capture the portion of credit risk covered by $IM_{\text{specific}}$ which ordinary $IM$ does not.

\medskip
\noindent Take $X_t = V_{t}-VM_{t}-IM_{g}$, where $IM_{g} = IM_{\text{general}}$ and  $\theta(X)$ to be the step function defined as follows
\[
\theta(X) =
  \begin{cases}
                                   0 & \text{if $X\leq 0$} \\
                                   1 & \text{if $X> 0$} \\
  \end{cases}
\]

\noindent Using the above definitions, we express Equation \ref{eq:CVA2} as

\begin{equation}
\label{eq:CVA5}
CVA_0^j = \mathbb{E}\left[ \int_{0}^{T} ( X_t-\alpha^j IM_{g} )\theta(X_t-\alpha^j IM_{g}) LGD^j_t PD^j_t DF_t dt   \right].
\end{equation}

\noindent Given that
\begin{multline}
( X_t-\alpha^j IM_{g} )\theta(X_t-\alpha^j IM_{g})\\  
= X_t \theta(X_t) + X_t[\theta(X_t-IM^j_{s})-\theta(X_t)] - IM^J_s \theta(X_t-IM^j_{s}), 
\end{multline}
\medskip

\noindent Equation \ref{eq:CVA5} can be further split into the following terms
\medskip
\begin{equation} \label{eq:rho1}
\begin{aligned}
\rho_1 &=  \mathbb{E}\left[ \int_{0}^{T} X_t \theta(X_t)LGD^j_t PD^j_t DF_t dt\right] \\ 
 &=  \mathbb{E}\left[ \int_{0}^{T}  (V_{t}-VM_{t}-IM_{g}) \theta( V_{t}-VM_{t}-IM_{g})
LGD^j_t PD^j_t DF_t dt\right],
\end{aligned}
\end{equation}

\medskip
\begin{equation} \label{eq:rho2}
\begin{aligned}
\rho_2 & = \mathbb{E}\left[ \int_{0}^{T} X_t [\theta(X_t - \alpha^j IM_{g})-\theta(X_t)] LGD^j_t PD^j_t DF_t dt\right],
\end{aligned}
\end{equation}

\medskip
\noindent where

\begin{multline}
\theta(X_t - \alpha^j IM_{g})-\theta(X_t)\\
 = (V_{t}-VM_{t}-IM_{g}) \theta( V_{t}-VM_{t}-IM_{g}-\alpha^j IM_{g}) - \theta( V_{t}-VM_{t}-IM_{g}),
\end{multline}

\medskip
\begin{equation}  \label{eq:rho3}
\begin{aligned}
\rho_3 & = - \mathbb{E}\left[ \int_{0}^{T} \alpha^j IM_{g} \theta(X_t - \alpha^j IM_{g}) LGD^j_t PD^j_t DF_t dt\right]  \\
& = \mathbb{E}\left[ \int_{0}^{T}  \alpha^j IM_{g} \theta( V_{t}-VM_{t}-IM_{g}- \alpha^j IM_{g}) LGD^j_t PD^j_t DF_t dt\right].
\end{aligned}
\end{equation}

\medskip
\noindent The term $\rho_1$ is Equation \ref{eq:CVA1} which is the CVA obtained with SIMM only. As has been explained in this paper, if this corresponds to the CVA of a badly rated counterparty, it may be difficult to hedge. The sum of terms $\rho_2$ and $\rho_3$ represent the portion of this CVA that specific IM helps us cover. Notice that $\theta(X_t - \alpha^j IM_{g})-\theta(X_t)$ is zero or negative, hence $\rho_2$ is always negative. What we are left with is Equation \ref{eq:CVA5} which is the CVA of a $AAA$ rated counterparty, which is much easier to hedge.

\section{Numerical Tests}
\label {sec:Numerical Tests}

\medskip
\noindent In this section we present the simulated $\alpha$ values and IM components ($IM_{\text{specific}}$ and $IM_{\text{add-on}}$) that were obtained for a collection of trades and counterparties. As we present these results, we make some notes and observations on how these quantities vary depending on trade characteristics.

\medskip
\noindent As explained in Section \ref{sec:IM in the Context of Risk Neutral Pricing} under the proposed framework, a AAA-rated counterparty will always have to post SIMM, while the rest will have to post $SIMM + \alpha^jSIMM$, where $\alpha^j$ depends on the counterparty’s rating and the financial instruments in the netting set. The values of $\alpha^j$ are computed by equating the CVA of the counterparty in question with the CVA of the AAA-counterparty, as expressed in Equation \ref{eq:CVA3}. The portfolios used for illustrative calculations are single-trade netting sets with at-the-money Interest Rate Swaps and European Swaptions, both of varying maturities. The European Swaptions were cash settled and always had an underlying swap of 5 years. Single-trade netting sets help identify how $\alpha$ varies according to the counterparty’s and trade’s properties. However, it must be noted that in a real-world setting, this calculation has to be done for netting sets composed of many trades, so the important inter-trade netting effects are accounted for.

\medskip
\noindent The methodology used to compute Dynamic Initial Margin (DIM) is based on Chebyshev Spectral Decomposition techniques, as it ensures exact DIM simulation per Monte Carlo path with very small computational cost \cite{Cheby}, ideal for optimisation problems such as this.

\medskip
\noindent The CVA values are computed using a Monte Carlo simulation with 10,000 paths. The root finding algorithm used to compute the value $\alpha$ for each netting set is the Newton-Ralphson root finding method.

\subsection{Dependency on credit rating}

\medskip
\noindent Figures \ref{bar_plot_IM_Swap_5} and \ref{bar_plot_IM_Swaption_ATM_5} show how the $IM_{\text{specific}}$ increases as the credit rating deteriorates. This happens in the examples with both Swaps and European Swaptions. This is clearly to be expected as higher probability of defaults mean higher CVA values. To compensate for higher CVA disparities, higher $IM_{\text{specific}}$ values are needed for Equation \ref{eq:CVA3} to hold.

\noindent

\begin{minipage}{\linewidth}
\makebox[\linewidth]{
  \includegraphics[keepaspectratio=true,scale=0.7]{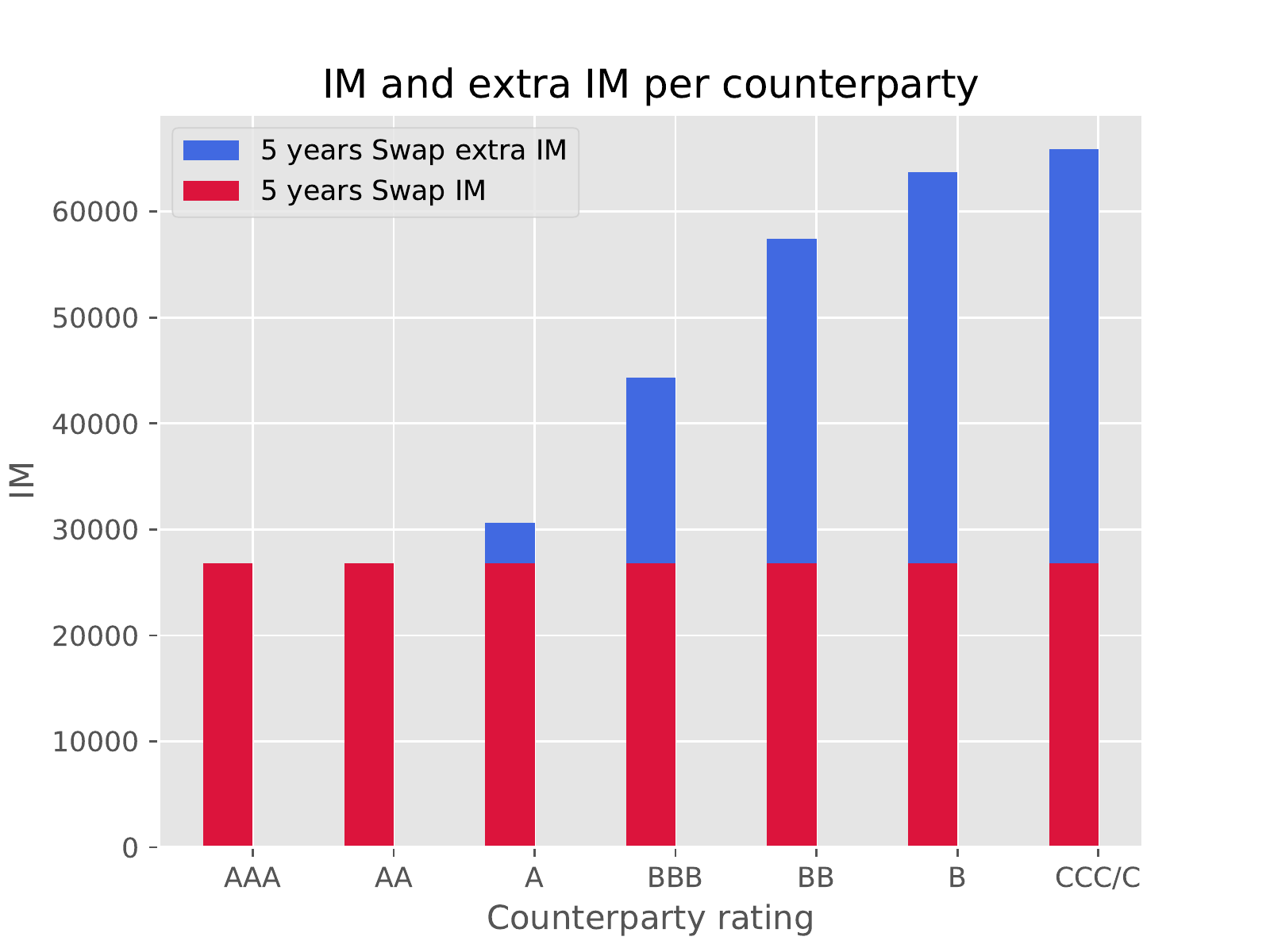}}
\captionof{figure}{} 
\label{bar_plot_IM_Swap_5}
\end{minipage}

\begin{minipage}{\linewidth}
\makebox[\linewidth]{
  \includegraphics[keepaspectratio=true,scale=0.7]{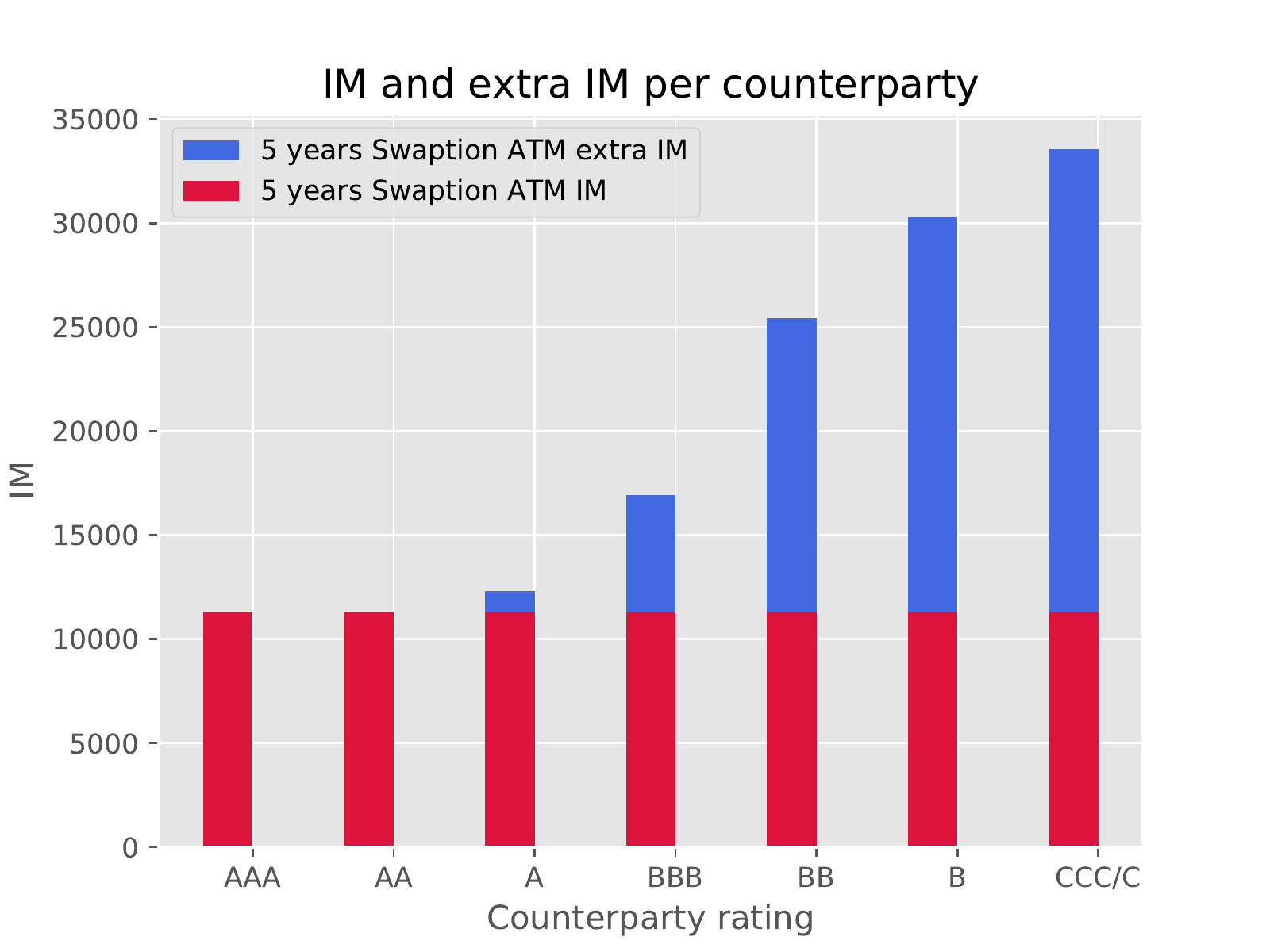}}
\captionof{figure}{} 
\label{bar_plot_IM_Swaption_ATM_5}
\end{minipage}

\medskip
\noindent An observation worth making is that the alpha values obtained were negative in some special cases (e.g. AA). This is as a result of the numerical noise of the Monte Carlo simulation and the noise in the profile of marginal default probabilities due to scarcity of data. As expected, these negative values are small, hence we floored alpha to zero. This is coherent with the fact that most solid financial institutions are at least AA rated, hence SIMM is a good value for IM to be posted between them.

\subsection{Dependency on portfolio maturity}

\medskip
\noindent As has been discussed above, as the credit rating decreases, the value of $IM_{\text{specific}}$ tends to increase. However, the rate at which this increase takes place can vary depending on the characteristics of the netting set. In figures \ref{bar_plot_IM_Swap_3_5} and \ref{bar_plot_IM_Swaption_ATM_3_5} we see that changing the maturity of the European Swaption drastically changed the rate at which the $IM_{\text{specific}}$ values increase across ratings. The $IM_{\text{specific}}$ demanded of the BBB counterparty is roughly twice the amount demanded of the AAA one; while the $IM_{\text{specific}}$ demanded of the CCC is more than twenty times higher the one demanded of the AAA counterparty. This contrasts with the Swap, where the increase of $IM_{\text{specific}}$ demanded across ratings increases much more linearly: the $IM_{\text{specific}}$ demanded of the BBB counterparty is roughly twice the one demanded of the AAA one, while the $IM_{\text{specific}}$ demanded of the CCC counterparty is only between 2 and 3 times higher.

\begin{minipage}{\linewidth}
\makebox[\linewidth]{
  \includegraphics[keepaspectratio=true,scale=0.8]{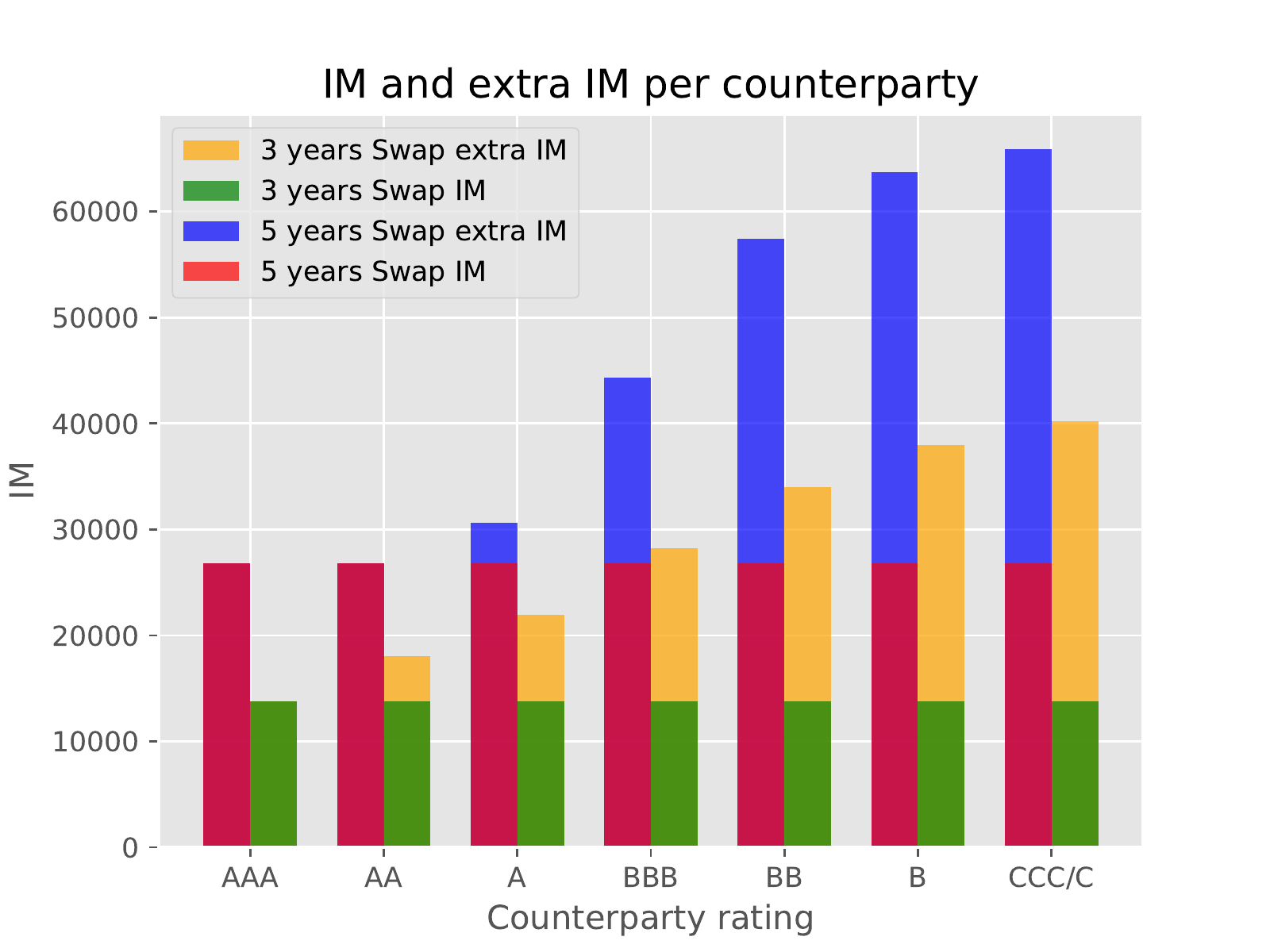}}
\captionof{figure}{} 
\label{bar_plot_IM_Swap_3_5}
\end{minipage}

\begin{minipage}{\linewidth}
\makebox[\linewidth]{
  \includegraphics[keepaspectratio=true,scale=0.8]{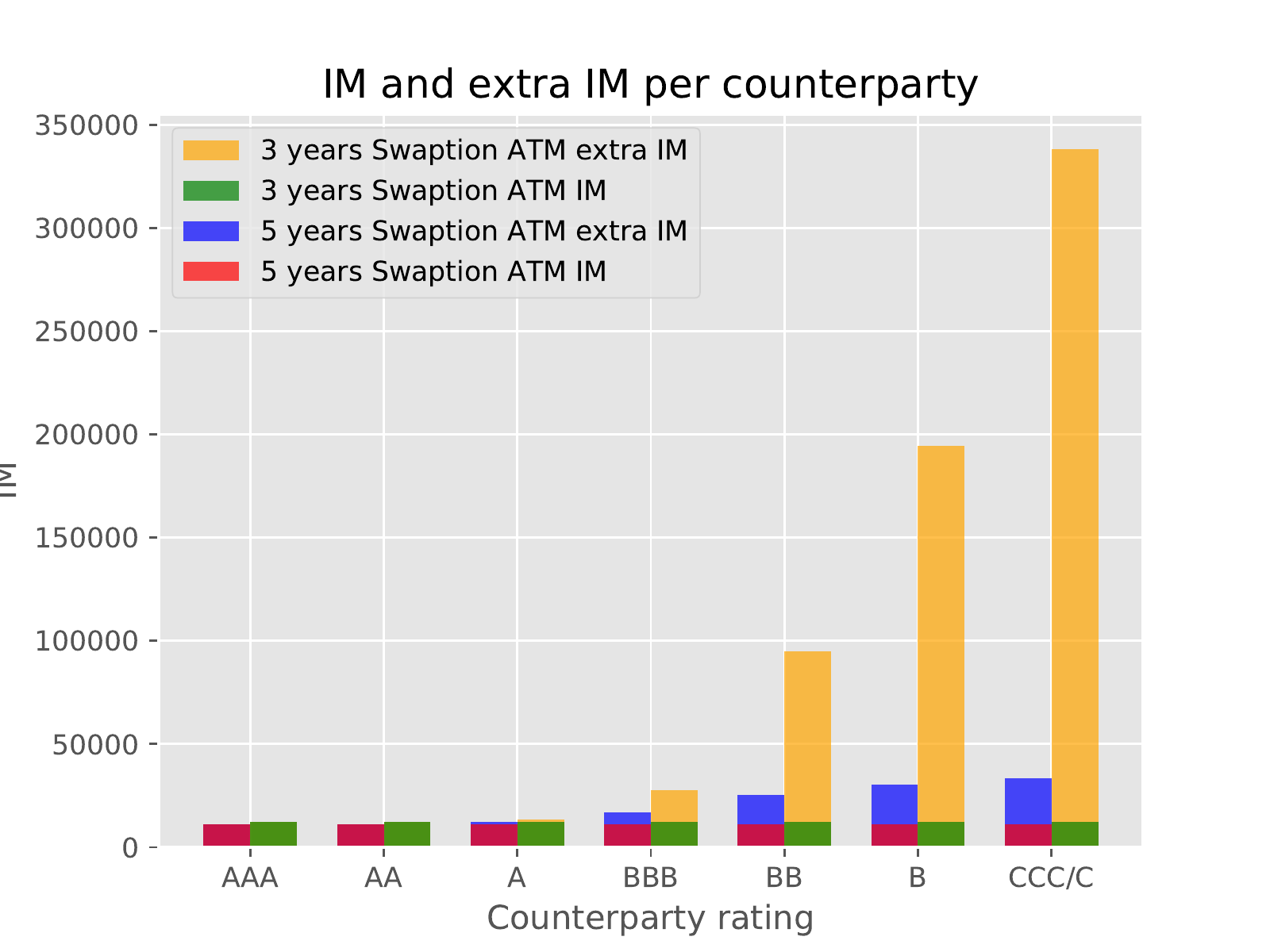}}
\captionof{figure}{} 
\label{bar_plot_IM_Swaption_ATM_3_5}
\end{minipage}

\medskip
\noindent Within the same trade type, even when the increase of $\alpha$ values is of roughly the same order of magnitude for trades of different maturities, the values themselves will be different. Take for example the Figure \ref{alphas_cptys_mats} which shows how the $\alpha$ values vary across credit ratings for swaps of different maturities. In general, as the maturity of the trade increases the $\alpha$ values tend to decrease for 
every counterparty.

\begin{minipage}{\linewidth}
\makebox[\linewidth]{
  \includegraphics[keepaspectratio=true,scale=0.8]{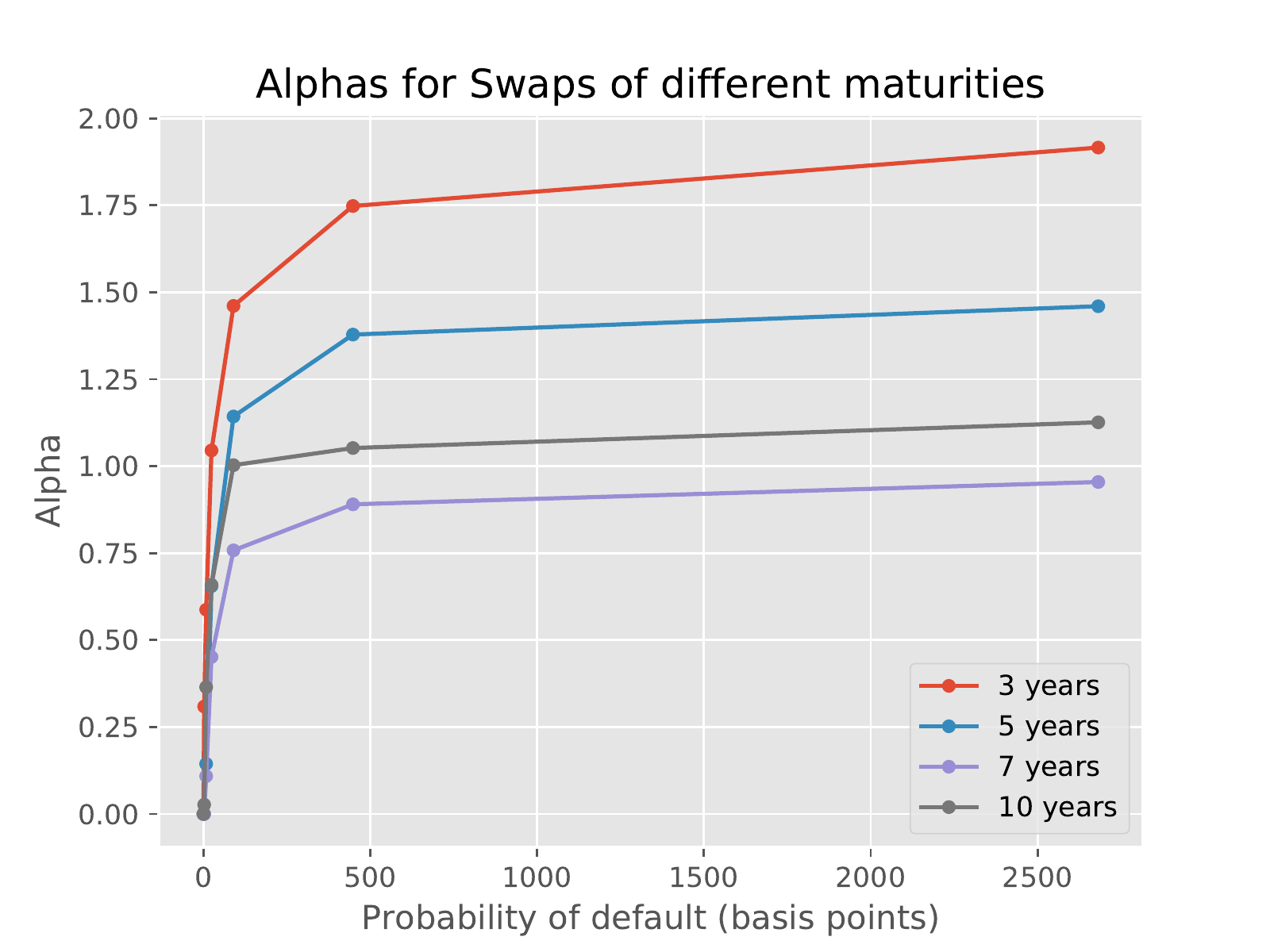}}
\captionof{figure}{} 
\label{alphas_cptys_mats}
\end{minipage}

\bigskip

\noindent Note also, as can be seen from tables \ref{tab:table2} and \ref{tab:table3}, that higher values of alpha do not mean higher levels of  $IM_{\text{specific}}$. In our example, counterparty CCC has an extra $IM_{\text{specific}}$ demand for a 10-year Swap which is more than twice that of the 3-year Swap; this despite $\alpha$ for the 10-year Swap being around half of that for the 3-year Swap. This of course is no surprise as today’s IM for both trades are different.

\medskip
\begin {table}[!htb]
\caption { IM values per rating.} \label{tab:table2}
\begin{center}
\begin{tabular}{|c|c|c|c|c|c|c|c|c|}
\hline 
\rule[-1ex]{0pt}{2.5ex} Maturity & IM & AAA & AA & A  & BBB & BB & B & CCC \\ 
\hline 
\rule[-1ex]{0pt}{2.5ex} 3 & 13,801 & 0 & 4,263 & 8,105 & 14,425 & 20,161 & 24,126 & 26,445 \\ 
\hline 
\rule[-1ex]{0pt}{2.5ex} 5 & 2,6789 & 0 & 0 & 3,859 & 17,547 & 30,615 & 36,924 & 39,108 \\ 
\hline 
\rule[-1ex]{0pt}{2.5ex} 7 & 35,912 & 0 & 0 & 3,906 & 16,208 & 27,220 & 31,971 & 34,227 \\ 
\hline 
\rule[-1ex]{0pt}{2.5ex} 10 & 51,832 & 0 & 1,380 & 18,900 & 34,143 & 51,980 & 54,536 & 58,363 \\ 
\hline 
\end{tabular} 
\end{center}
\end{table}

\begin {table}[!htb]
\caption { Alpha values per rating } \label{tab:table3}
\begin{center}
\begin{tabular}{|c|c|c|c|c|c|c|c|}
\hline 
\rule[-1ex]{0pt}{2.5ex} Maturity  & AAA & AA & A  & BBB & BB & B & CCC \\ 
\hline 
\rule[-1ex]{0pt}{2.5ex} 3  & 0 & 0.308907 & 0.58726 & 1.045221 & 1.460771 & 1.748091 & 1.916069 \\ 
\hline 
\rule[-1ex]{0pt}{2.5ex} 5 & 0 & 0 & 0.144077 & 0.655038 & 1.142833 & 1.378342 & 1.459857 \\ 
\hline 
\rule[-1ex]{0pt}{2.5ex} 7  & 0 & 0 & 0.108764 & 0.451318 & 0.757964 & 0.890267 & 0.954463 \\ 
\hline 
\rule[-1ex]{0pt}{2.5ex} 10  & 0 & 0.026635 & 0.364651 & 0.658728 & 1.002851 & 1.05217 & 1.126013 \\ 
\hline 
\end{tabular} 
\end{center}
\end{table}

\pagebreak
\noindent There is nothing in the examples presented that cannot be generalised to more complex real-world netting sets. A financial institution should be able to compute, given a netting set, and a counterparty, the proportion ($\mathbf{\alpha}$ value) of SIMM that needs to be added to the IM required of the counterparty to turn the netting set into a AAA-quality one. As in the CVA world, in which it is the incremental CVA that is priced into the trade, in this IM world it is the incremental IM that is requested from the counterparty for the execution of the trade. Just as with CVA, this should be computed for every incoming trade and adjusted on ever existing netting set on a daily basis, or as often as IM is computed.

\medskip
\noindent It is important to note that computing the $\alpha$ values corresponding to a given trade (or set of trades) and different counterparties is computationally very demanding. One must be able to run a Monte Carlo simulation of market-to-market values (to compute VM) and SIMM (or any other IM value). In this paper we were able to do such computations in a single and ordinary PC using Chebyshev Spectral Decomposition methods as proposed in \cite{Cheby}.

\section{Conclusion}
\label {sec:Conclusion}

\medskip
\noindent In this paper we have introduced an upgraded methodology for Initial Margin (Specific Initial Margin) that depends on the credit rating of the counterparty. As opposed to the current CVA amounts which are negligible, we have shown that the extra IM required is significant specially for the lowest rated counterparties.

\medskip
\noindent 
The new Specific Initial Margin is composed of two quantities: the IM as computed today in the industry (e.g. SIMM) plus an add-on, which depends on the counterparties’ credit rating. This second component is computed so that under a CVA framework, the CVA of the corresponding netting set is reduced to that of a AAA-rated counterparty (as described in \ref{sec:Specific Initial Margin}). Note that other risk metrics could be used (e.g. peak PFE); however, we think CVA is optimal given its widespread use in the industry and its sound pricing risk-neutral foundations.

\medskip
\noindent  
This new type of IM has the following advantages. First, by reducing the CVA of the counterparty via extra IM we adjust the counterparties effective credit rating. This is convenient given that CVA, in the real world, can fail to provide the protection for which it was designed. By transferring the CVA value to an IM add-on, we take advantage of the mechanisms in place within the industry that ensures the posting and use of IM amounts when needed, hence protecting against CCR more effectively \footnote{In the case of bilateral IM, Basel’s directives require it to be posted on a third-party segregated account with no rehypothecation option. This ensures that the IM amount will be available for the surviving party in the case of a default.}. Moreover, in this way, everything is in compliance with the risk-neutral pricing assumptions.

\medskip
\noindent Lastly, given the way $IM_{\text{specific}}$ is defined, the lower the credit rating of a counterparty, the higher $IM_{\text{specific}}$ will be. This makes total sense from a financial standpoint. In this paper we proposed a methodology to compute $IM_{\text{specific}}$, both for total and trade-incremental values, in an effective and sound manner.


\medskip
\medskip



\begin{thebibliography}{1}
\bibitem{BCBS} Basel Committee on Banking Supervision and International Organization of Securities Commissions (2015). {\em Margin Requirements for Non-Centrally Cleared Derivatives}. D317 March 2015.

\bibitem{Pykhtin} Pykhtin, M. (2009),\emph{ Modelling Credit Exposure for Collateralised Counterparties}. Journal of Credit Risk, 5(4), pages 3-27.

\bibitem{Andersen-MPOR} Andersen, L and Pykhtin, M and Sokol, A (2017),\emph{Rethinking the Margin Period of Risk}. Journal of Credit Risk 13(1), 1-45.

\bibitem{Andersen-Exposure} Andersen, L and Pykhtin, M and Sokol, A. \emph{Credit Exposure in the Presence of Initial Margin}. (July 22, 2016). Available at SSRN: https://ssrn.com/abstract=2806156 

\bibitem{Andersen-IM} Andersen, L. and M. Pykhtin, and A. Sokol (2017). \emph{Does Initial Margin Eliminate Counterparty Risk?}. Risk Magazine, May 2017.

\bibitem{Brigo}	Brigo, D., Morini, M., Pallavicini, A. \emph{Counterparty Credit Risk, Collateral and Funding}. Wiley, 2013.

\bibitem{ISDA-SIMM}	ISDA SIMM – Methodology, version R.1.3, Effective Date: April 1, 2017

\bibitem{Burgard}	C. Burgard, M. Kjaer.\emph{ Funding Costs, Funding Strategies}. Risk, 82-87, Dec 2013. 

\bibitem{Cheby}	Ruiz I., Zeron M. \emph{Dynamic Initial Margin via Chebyshev Spectral Decomposition}. arXiv:1808.08221

\bibitem{Gregory}	Gregory, J. \emph{ The Impact of Initial Margin} (May 16, 2016). Available at SSRN: https://ssrn.com/abstract=2790227 



\end{thebibliography}
\end{document}